\documentclass[aps,epsfig,twocolumn,showpacs]{revtex4-1}
\usepackage{amssymb}
\usepackage{graphics}
\usepackage{graphicx}
\usepackage{epsfig}
\usepackage{epstopdf}
\begin{document}
\title{Differential stability of DNA based on salt concentration}
\author{Arghya Maity}
\author{Amar Singh}
\thanks{A. Maity and A. Singh contributed equally to this work.}
\author{Navin Singh}
\affiliation{Department of Physics, Birla Institute of Technology \& Science, Pilani - 333 031, Rajasthan, India}

\begin{abstract}
Intracellular positive ions neutralise  negative charges on the phosphates of a DNA strand conferring greater strength on the hydrogen bonds that connect complementary strands into a double helix and so confer enhanced stability. Beyond a certain value of salt concentration, the  DNA molecule displays a unstable nature {\it in vivo} as well as {\it in vitro}. We consider a wide range of salt concentrations and study the stability of the DNA double helix using a statistical model. Through numerical calculations we attempt to explain the different behaviour exhibited by DNA molecules in this range. We compare our results with experimental data and find a close agreement.


\end{abstract}
\maketitle

\section{Introduction} 
Intracellular sodium triggers a cell to progress to cell division \cite{cone1985}. In many biological activities like creation of daughter cells, salt enacts a pivotal role.  It is also a known fact that the two strands of DNA carry negative charge due to their phosphate groups ($PO_4^-$) \cite{watson}. The two strands are connected with each other through the hydrogen bonds between the bases on the opposite strands. However, the stability of the DNA double helix also depends on the charge neutrality of the system. To counter the negative charges on the strands, the cellular environment requires some positive charges in order to nullify the repulsion between negative charges. Such positive counter ions derive from  salts present in the body. 
The primary salts are $NaCl$ and $MgCl_2$ which on reaction with water release $Na^+$ and $Mg^{2+}$, while the chlorine rearranges itself in the form of hydrogen chloride ($HCl$). These cations act as  electrostatic screening agents by creating an attractive force between the positive ions  \cite{manning,manning1} and the negative charges on the strands. The stability of the DNA helix is then governed by the balance between the attractive and repulsive forces exist in the system. These forces exist in the form of hydrogen bonding between the bases on the opposite strands, the stacking interaction along the strands, 
the repulsion between same kind of charges and attraction between opposite kind of charges. Various experiments followed by theoretical investigations on the role of salt in screening the repulsive forces in DNA revealed many interesting results. Not only the stability of DNA molecule \cite{amar2013, amar2013phyexp,krueger} but effects of salt on the $B-A$ transition, and on the condensation of the DNA molecule have been studied in detail by several researchers \cite{hormen01,hormen02}. In all these studies the stability of DNA has been studied under low or moderate (0.1-1.0 M) concentrations of salt. Several experiments, conducted by the groups of Owczarzy \cite{owc, owc1} and SantaLucia \cite{santa, santa1} revealed that short as well long DNA molecules 
become more stable as the content of cations in the solution increases, with the stability of the structure being determined by calculating the melting temperature ($T_m$) of the molecule. In all these experiments, the melting temperature was found to have a logarithmic dependence on the amount of cations present in the solution. The theoretical work following these experiments attempted to explain this logarithmic dependence, based either on a semi-empirical formula \cite{Chaurasiya,krueger,nik} or on statistical models \cite{amar2013,amar2013phyexp,amb,amb1,weber2015}. 
The counterion theory proposed by Manning explains the stability of the DNA molecule due to the presence of salt in the solution \cite{manning,manning1}. The stretching and unzipping behaviour of the DNA molecule in the presence of cations have also been discussed by several groups \cite{dong, huguet, amar2013}. These results show that the mechanical stability of the DNA molecule also increases with the concentration of cations. 

In another set of experiments \cite{sebstian_jacs96, khimji} that were executed at relatively high salt concentration, some strikingly different behaviour in the DNA molecule was observed. These experiments found that in this range of salt concentrations the stability of DNA gets shattered. Interestingly, in a similar set of experiments, 
several research groups tested
the condensation process and melting profile in DNA under different ethanol concentrations \cite{hormen01,pikur}. One of the findings of these experiments revealed that at a relatively high concentration of ethanol, the melting temperature of the system increases. The role of cations {\it in vivo} is manifold. In a recent review, Bose {\it et al} \cite{bose2015} showed that abundant cations inside the cell as well as outside the cell play important roles in sustaining cancer cells and at the same time in the decay of immune cells. Cancer cells possess a 
significant electrical character\cite{cureJC} compared to normal somatic cells. Cancer cells become more electrostatic \cite{cone1985} during oncogenesis.

Keeping in view the important aspect of  salt in the cell, we propose a theoretical description of the stability of DNA molecules at higher salt concentration. We consider one of the DNA sequences studied by Khimji {\it et al.} \cite{khimji}, who studied the DNA duplex stability in crowded polyanion solutions. For the current investigations we adopt the standard Peyrard Bishop Dauxois (PBD) model \cite{pb, pb1} and modify the potentials appearing in the model. 
In our earlier work, we showed that the PBD model has enough detail to explain the thermal as well as 
mechanical stability of DNA molecules at lower strengths of salt and with varying salt concentration \cite{amar2013,amar2015}. Our presentation in the current manuscript is organized as follows. In section \ref{model}, we discuss the PBD model and the suitable modifications we did to incorporate the high concentration of cations. How the stability of DNA in thermal as well as in force ensembles varies in a wide  range of salt concentrations in solution is discussed in section \ref{result}. We finally conclude our results in section \ref{concl}.

\section{Model}
\label{model}

We adopt the PBD model to study the response of DNA  in different ensembles. We use the linear version of the PBD model. Although this version simplifies the helicoidal geometry of the molecule, it has enough details to delineate the stability of the molecule in thermal as well as in force ensembles \cite{barbi,cocco,kumar2010}. The model represents the hydrogen bonding between the bases in a pair through the Morse potential while the stacking interaction is represented by an anharmonic potential. 
The Hamiltonian of the system is,
\begin{equation}
\label{eq1}
H = \sum_{i=1}^{N}\left[\frac{p_i^2}{2m} + V_m(y_i) + V_{sol}(y_i)\right] 
+ \sum_{i=1}^{N-1} V_s(y_i,y_{i+1})
\end{equation}
where $(p_i = m${\it \.y$_i$}) represents the momentum of a base pair where $m$ is the mass of a base pair and $y_i$ represents the stretching of the $n^{th}$ base pair. The stacking interaction between the nearest neighbours is represented by an anharmonic term, $V_s(y_i,y_{i+1})$, which is,
\begin{equation}
\label{eq2}
V_s(y_i,y_{i+1})= \frac{k}{2}(y_i-y_{i+1})^2[1+\rho\exp\{-b(y_i+y_{i+1})\}].
\end{equation}
Here $k$ represents the single strand elasticity. The anharmonicity in the strand elasticity is taken care of by the term $\rho$ while $b$ represents the range of  anharmonicity. For our studies we tuned the model parameters and found the values of $k = 0.01 \; {\rm ev/\AA^2}$, $\rho = 1.0,\; b = 0.035$ \AA\ as suitable values. The potential $V_m(y_i)$ describes the interaction between the two bases in a pair \cite{pb,pb1}
\hspace{0.2in}
\begin{equation}
\label{eq3}
V_m(y_i)= D_i({e}^{-a_iy_i}-1)^2
\end{equation}
where $D_i$ is the equivalence of the dissociation energy of a pair. The constant, $a_i$, represents the inverse of the width of the potential well. These two parameters have a crucial role in  DNA denaturation. The dissociation energy is a representation of the hydrogen bond energy that binds the $A-T$ and $G-C$. It is a known fact that the bond strength of these two pairs are not the same but are in an approximate ratio of 1.25-1.5 as the $GC$ pairs have three while $AT$ pairs have
two hydrogen bonds \cite{ares, nik, weber, zoli, zoli1,zoli2, ffalo, ffalo1, ffalo2, frank, macedo}. The last term in the Hamiltonian is the solvent term $V_{sol}(y)$. 
This is an additional term which simulates the formation of H-bonds with the solvent, 
once the hydrogen bonds are stretched by more than their equilibrium values. This is expressed as\cite{zhang1995, amar2013}
\begin{equation}
\label{eq6}
V_{\rm sol}(y_i)=-\frac{1}{4} D_i [\tanh(\gamma y_i) -1]
\end{equation}
where $\gamma$ is the solvent interaction factor and it reduces the height of the barrier appearing in the potential \cite{zhang1995}.

\begin{figure}[h]
\includegraphics[height=2.0in,width=3.0in]{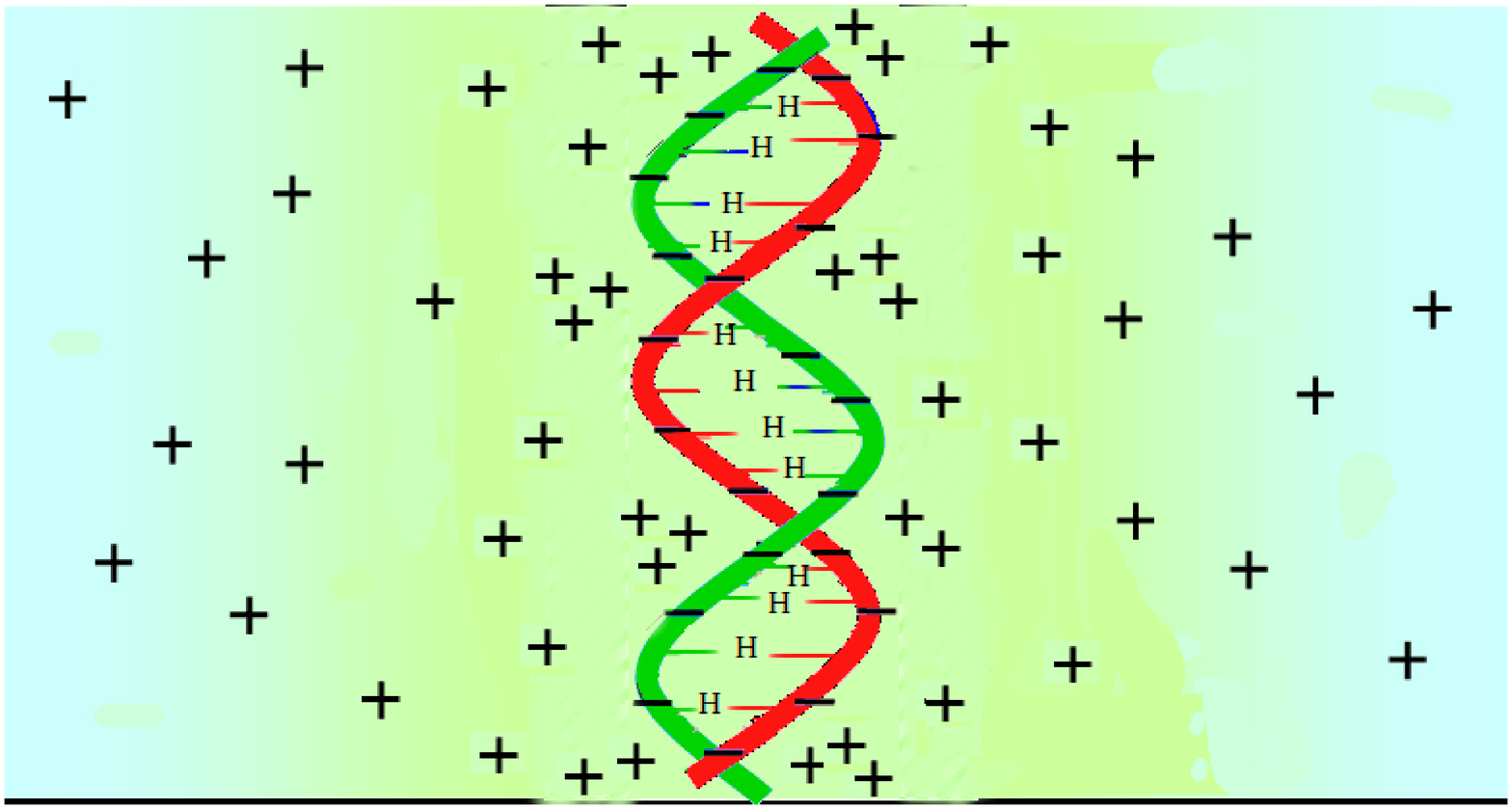} \hspace{.5in}
\caption{\label{fig01}The schematic representation of cations and DNA. The figure shows the probable distribution of cations around the two negative strands which are standing around 20\u{A} apart and the cations trying to nullify the repulsion between the two strands and possible distribution of excess cations that interact among themselves. 
}
\end{figure}

The concentration of salt that stabilizes the DNA molecule can be incorporated through the on-site potential term \cite{dong,amar2013} in the Hamiltonian. The presence of cations around the DNA molecule are shown, schematically, in Fig. \ref{fig01}.
It is a known fact that the cations are required to reduce the repulsion between and along the DNA strands so that the DNA is stabilised in its double stranded configuration. However, there is a critical concentration of cations in the solution below which these cations act as  shielding agents. In case the number of cations in the solution crosses this critical number, the repulsive forces in the system dominate over the attractive forces which further shatter the stability of the dsDNA. 
The overall activities in the surroundings of DNA molecule at higher concentration might be complicated to analyse. However, as an intuitive guess, one can anticipate the behaviour of cations as Coloumbic particles. We are helped by the basics of electrostatics here. From elementary knowledge we know that for a monopole the potential scales as $\frac{1}{r}$, where $r$ is the distance between the charges. Similarly for dipole it scales as $\frac{1}{r^2}$, for quadrupole $\frac{1}{r^3}$ and for octupole as $\frac{1}{r^4}$  etc \cite{griffith}. We extend this concept and introduce a scaling term that represents the repulsive contribution in the complete potential. 
The cation's concentration, $C$ around the DNA molecule can be mapped with the inter particle distance $r$. The logic behind this proposal lies in the fact that when $C$ increases, $r$ will decrease and vice versa. In the PBD model, the prime contributor to the stability of the molecule is the potential depth of the Morse potential ($D_i$). Thus, we adopt this parameter ($D_i$) as a function of the salt concentration of the solution. Considering all of the above forces that are responsible for the stability of DNA molecules, we express the parameter, $D_i$, as (in terms of salt concentration),
\begin{equation}
\label{eq4}
D_i = D_0\left[1+\lambda_1\ln \left(\frac{C}{C_0}\right) + \lambda_2\ln^2 \left(\frac{C}{C_0}\right) + \chi \left(\frac{C_0}{C^t}\right)\right]
\end{equation}
where the first three terms are taken from our previous work \cite{amar2013} which show the logarithmic dependence of $T_m$ on salt concentration. Here we make $C$ as a dimensionless quantity by choosing the reference value $C_0 = 1.0$ \cite{amar2013}. The fourth term is added which takes care of the behaviour of cations in the surrounding of DNA molecules at high salt concentrations. This is a pure Coloumbic term. A point to note is that $t$ is an important parameter. 
Since the nature of cations and their exact distribution is not known, how the Coloumbic interactions modify the interaction between the two strands is not precisely known. At the same time, it is also a fact that cations shield the repulsion between the negative ions. We are not sure about the nature of the {\it pole} that might be existing due to presence of these cations. The best we can think of is that they obey some power law. We propose that the  potential depth will scale with the concentration of salt as $C^{-t}$. 
This term will have a proportionality factor $\chi$ which can be found by comparing the theoretical results with the experimental findings. At lower values of $C$, the logarithmic term  will dominate, so the molecule will be more stable with increasing values of $C$. For higher concentration the interaction between the cations will dominate. For our calculations we found the best match for $t = 0.01$ and $\chi = 1.2$. The other potential parameters are $D_0 = 0.043$ eV, $\lambda_1 = 0.01$, $\lambda_2 = -0.011$, $a_{AT}=4.2\; {\rm\AA}^{-1}$, $a_{GC} = 6.3 \; {\rm \AA}^{-1}$.  
We take $\gamma = 1.0 \; {\rm \AA}^{-1}$ in this problem as a suitable choice for our calculations.

The partition function can be written with the complete Hamiltonian of 
the system like this,
\begin{equation}
\label{eq7}
Z= \int \int \prod_{i=1}^{N} \exp [-\beta H(y_i,y_{i+1},p_i)] dy_i dp_i
\end{equation}
where $\beta=\frac{1}{k_B T}$. The momentum part can be easily integrated and is equal to $(2\pi mk_BT)^{N/2}$. 
The configurational partition function is defined as,
\begin{equation}
\label{eq8}
Z_c= \int \prod_{i=1}^{N} K(y_i, y_{i+1}) dy_i
\end{equation}
For the homogeneous chain, one can evaluate the partition function by transfer integral (TI) method by applying the periodic boundary condition \cite{chen}. In the case of a heterogeneous chain, with an open boundary, the onfigurational part of the partition function can be integrated numerically with the help of the matrix multiplication method \cite{chen,ns2001,erp,ns2011}. The important part of this integration is the selection of proper cut-offs for the integral appearing in Eq.6 to avoid the divergence of the partition function.
The method to identify the proper cut-off has been discussed by several groups \cite{chen,pbd95,erp}. In the calculations by T.S. van Erp {\it et al} it was shown that the upper cut-off should be $\approx$ 50 \AA\ with  our model parameters at $T = 300$ K while the lower cut-off is -0.3 \AA. In the earlier work by Dauxois and Peyrard it was shown that the $T_m$ converges rapidly with the upper limit of integration \cite{pbd95}. In this work the equation for the partition function for an infinite homogeneous chain was solved using the TI method. For a short chain, we calculated $T_m$ for different values of 
upper cut-offs and found that an upper limit of 200 \AA\ is sufficient to avoid the divergence of the partition function. Thus the 
configurational space for our calculations extends from -5 \AA\ to 200 \AA. Once the  limit of integration has been chosen, the task is reduced to discretize the space to evaluate the integral numerically. The space is discretized using the Gaussian quadrature formula. In our previous studies \cite{ns2005}, we observed that in order to get precise value for melting temperature ($T_m$) one has to choose  large grid points. We found that 900 is a sufficient number for this purpose. As all matrices in Eq. \ref{eq8} are identical in nature the multiplication is done very efficiently.
The thermodynamic quantities of interest can be calculated by evaluating the Helmholtz 
free energy of the system. The free energy per base pair is,
\begin{equation}
\label{eq9}
f(T) = -\frac{1}{2\beta}\ln\left(\frac{2\pi m}{\beta}\right) - 
\frac{1}{N\beta}\ln Z_c; \qquad\qquad \beta = \frac{1}{k_BT}.
\end{equation}
The thermodynamic quantities like specific heat ($C_v$) as a function of temperature or the applied force can be evaluated by taking the second derivative of the free energy. The peak in the specific heat corresponds to the melting temperature or the critical force of the system.

Other quantities like the average fraction $\theta(= 1 - \phi)$ of bonded (or open) base pairs can be calculated by introducing the dsDNA ensemble(dsDNA) \cite{erp} or using the phenomenological approach \cite{campa,ns2001}. 
In general, the $\theta$ is defined as,
\begin{equation}
\label{eq10}
\theta = \theta_{\rm ext}\theta_{\rm int} 
\end{equation}
$\theta_{\rm ext}$ is the average fraction of strands forming duplexes, while $\theta_{\rm int}$ is the average fraction of unbroken bonds in the duplexes. The opening of long and short chains are completely different. For long chains, when the fraction of open base pairs, $\phi(1-\theta)$, goes practically from 0 to 1 at the melting transition, the two strands are not yet completely separated. At this point, a majority of the bonds are in a broken state and the dsDNA is denatured. However, few bonds are still in an intact state, preventing the two strands separating from each other. Only at high temperatures there is a real separation. Therefore for very long chains the double strand
is always a single macromolecule through the transition, thus one can calculate the fraction of intact or broken base pairs only. While for short chains, the process of single bond disruption and strand dissociation tends to happen in the same temperature range. Thus, the computation of both $\theta_{\rm int}$ and $\theta_{\rm ext}$ is essential \cite{campa}. 

\section{Results}
\label{result}

\begin{figure}[h]
\includegraphics[height=2.3in,width=2.6in]{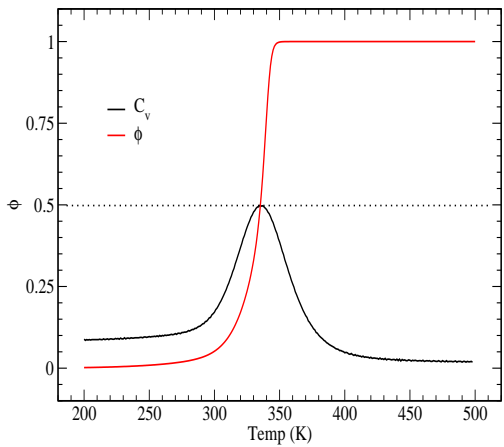}
\caption{\label{fig02}  
The specific heat as a function of temperature for the 12 base pair chain for the model parameters $k = 0.01\; {\rm eV/\AA^2}, \; D_0 = 0.043 \; {\rm eV}, \;\lambda_1 = 0.01,\; \lambda_2 = -0.011, \; t = 0.01,\; \chi = 1.2$. 
The figure is plotted for $C = 0.5$ M. The peak corresponds to the melting temperature.}
\end{figure}

\begin{figure}[h]
\includegraphics[height=2.3in,width=2.8in]{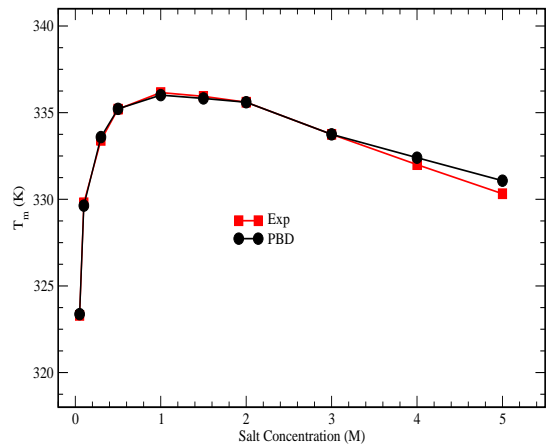} \hspace{.1in}
\caption{\label{fig03}
The temperature-salt phase diagrams showing the variation of $T_m$ with salt concentration.
The comparison of our results with the experimental results of \cite{khimji} }
\end{figure}

First we discuss the thermal denaturation of the DNA molecule. To find the thermal stability of the molecule we calculated the melting temperature with the help of Eq. \ref{eq9}\&\ref{eq10}. Here we discus the stability of the short DNA sequences that are used by Khimji {\it et al} \cite{khimji}. We consider the sequence $5'-TCACAGATGCGT-3'$ for our studies. The peak in the specific heat and in the differential melting curve indicate the melting temperature of the molecule (see Fig. \ref{fig02}). We calculate the melting temperature of DNA for values of salt ranging from 0.05 to 5.0. The resultant phase diagram is shown in Fig.\ref{fig03}.
From the figure we find that the first three terms in Eq.(4) are responsible for the behaviour of the molecule in the low salt region, between 0.01-1.0. Once the concentration is above 1.0 the Coloumbic repulsion among the cations increases. This leads to the instability in the molecule. As discussed earlier, the parameter $t$ plays an important role here. We calculate the phase diagram for different values of $t$ and find the best match for $t = 0.01$ for the 
experimental results of Khimji \cite{khimji}. The maximum deviation is at $C = 5.0$ which is about 0.8 K. Thus, we can say that our results are in close match with the experimental findings of \cite{khimji}.  

\begin{figure}[h]
\includegraphics[height=2.3in,width=2.8in]{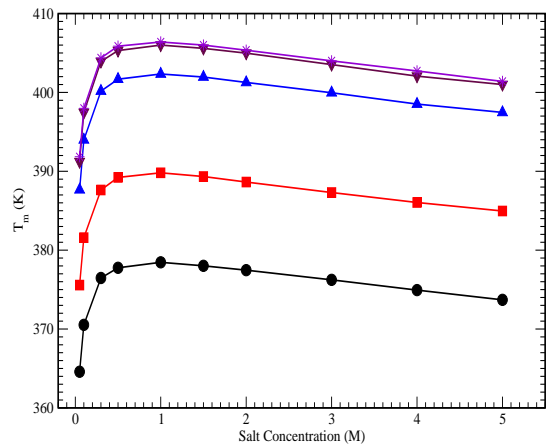} \hspace{.1in}
\caption{\label{fig04}
The phase diagram for different chain lengths in order to validate the results for infinite chain. The lengths are 36, 60, 156, 252, 276 bps.}
\end{figure}

In order to validate our results for long chains, we increased the length of the molecule by repeating the same sequence. From Fig. \ref{fig04} we find the behaviour of phase diagram is the same with increasing lengths of DNA molecule. However, as the chain length increases the phase diagram for different lengths comes closer. This indicates the limit of infinite chain length. We do not have the experimental data for infinite chains at this moment to compare our results.

It is important to note here the difference in the nature of the slope at lower and higher concentrations. Experimental 
as well as theoretical findings clearly indicate that the stability rate is much sharper than the instability rate. 
The prime reason behind this difference might be the effect of excluded volume at higher concentrations. As the concentration of the salt increases the number of cations increase correspondingly and therefore the volume available for the movement will lessen. The cations have charge as well as size and so occupy the flexible volume (they crowd the DNA). 
At higher concentrations although there is a instability due to repulsive forces, there is a stability due to excluded volume. While the movement of cations at lower concentration is faster, at high concentration their movement might be hampered.  

After evaluating the stability of DNA in a thermal ensemble we calculated the stability of the molecule in a force ensemble for the above mentioned range of salt concentration in the solution. It is a known fact that the replication process is initiated by the force exerted by DNA polymerase on a segment of DNA chain. Replication starts at the replication origin \cite{branzei,kaufmann} and the replication fork propagates bidirectionally. Mathematically one can model the replication as the force applied on an end of the DNA chain \cite{garima,somen}. The physics of opening the chain in these ensembles is completely different \cite{hatch,huguet}.Whereas in thermal denaturation, the opening is 
due to an increase in the entropy of the system, for a mechanically stretched DNA chain the opening is enthalpic. The modified Hamiltonian for the DNA that is pulled mechanically from an end is,
\begin{equation}
\label{eq11}
H = \sum_{i=1}^N\left[\frac{p_i^2}{2m}+ V_m(y_i) + V_{sol}(y_i)\right] +
\sum_{i=1}^{N-1}\left[W_S(y_i,y_{i+1})\right] -F\cdot y_1
\end{equation}
where the force $F$ is applied on the $1^{st}$ base pair \cite{ns2005}.
The thermodynamic quantities of interest, from the modified Hamiltonian, can be calculated using Eq. \ref{eq8} \& \ref{eq9}. Here we consider an infinite chain of 300 base pairs which is generated by repetition of the 12 base pair chain \cite{khimji}. {\it In vitro,} experiments on DNA unzipping are executed either at constant displacement or at varying loading rates. Although in both the set-ups, the microscopic and dynamic behavior of unzipping are different, the critical force comes out to be the same for an infinite chain (or in thermodynamic limit), {\it i.e.} for $\lambda$-phase DNA. For theoretical investigation on force induced unzipping of DNA, the chain length of the molecule should be large enough to consider the chain as infinite. We calculated the melting temperature, 
$T_m$ for different chain lengths (in increasing order) and found that a length of about 300 base pairs is
sufficient to consider the chain as an infinite chain \cite{amar_macro}. All the base pairs of dsDNA kept in a thermal bath share equal amount of energy while in the case when the molecule is pulled from an end there is a differential distribution of the applied force (from the pulling point to the other end of the chain). 
\begin{figure}[]
\includegraphics[height=2.3in,width=2.8in]{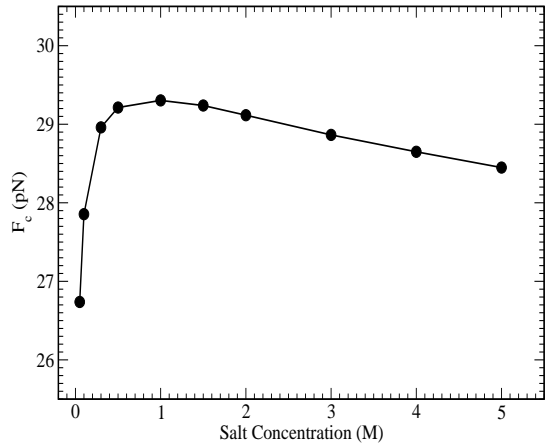}
\caption{\label{fig05} The critical force-salt phase diagram showing the variation of $F_c$ with salt concentration. 
As there is no experimental data is available to compare, we show only the calculated points on the graph.}
\end{figure}
The phase diagram for the mechanically pulled DNA shows that the  force required to open the chain  decreases at 
higher salt concentrations in the solution, as shown, Fig.\ref{fig05}. Since the molecule itself is unstable at higher 
salt concentrations, the force required is less. Unfortunately, we do not have any experimental data for force induced 
unzipping at higher salt values. We hope that some experiments may be performed to validate these results.

\section{Conclusion}
\label{concl}

In the current work, we propose a theoretical description of the behaviour of DNA molecules at relatively high concentrations of salt. There are several biological events where the concentration of cations increases in the cell. For example, in somatic cells the concentrations of sodium ions substantially increase during activation to become cancer cells \cite{kaufmann, cone1985}. From most of the previous studies it was concluded that these cations contribute to the stability of the DNA in the cell \cite{amar2015,weber}. These cations act as shielding agents for the negatively charged strands of DNA. However, there is a range up to which they contribute to the stability of the molecule. When the concentration exceeds this range, their role in stability may be shattered \cite{khimji}. Our work is an attempt to explain the possible mechanism that might be taking place at higher concentrations of salt in the solution. 
The volume available in the cell is fixed and there is a limit up to which these cations contribute to the overall charge neutrality of the system. A greater number of cations in a fixed volume may disrupt the balance between the positive and negative charge forces. Through the schematic diagram and by calculating the free energy as a function of temperature and force, we explain the cause of the instability of the DNA molecule at higher concentrations. Our results are in very good agreement with the experimental findings \cite{khimji}. However, due to unavailability of results for infinite chains we can not compare our results for force induced unzipping of DNA molecules at high salt concentration. It would be interesting to calculate the kinetics of duplex unzipping at higher salt concentration. The movement of cations at higher salt concentrations might also be interesting to analyse.

\section*{Acknowledgement}
We are thankful to S. Kumar and Y. Singh, Department of Physics, Banaras Hindu University, India,for useful discussions. We acknowledge the financial support provided by Department of Science and Technology, New Delhi under the project grant (SB/S2/CMP-064/2013). A.S. would like to thank UGC, India for the financial support through BSR fellowship.


\end{document}